# Human as Real-Time Sensors of Social and Physical Events:
## A Case Study of Twitter and Sports Games

Siqi Zhao and Lin Zhong
Dept. of Electrical & Computer Engineering
Rice University, Houston TX
{siqi.zhao, lzhong}@rice.edu

Jehan Wickramasuriya and Venu Vasudevan
Applied Research
Motorola Mobility, Libertyville IL
{jehan, venu.vasudevan}@motorola.com

## Abstract
In this work, we study how Twitter can be used as a sensor to detect frequent and diverse social and physical events in *real-time*. We devise efficient data collection and event recognition solutions that work despite various limits on free access to Twitter data. We describe a web service implementation of our solution and report our experience with the 2010-2011 US National Football League (NFL) games. The service was able to recognize NFL game events within 40 seconds and with accuracy up to 90%. This capability will be very useful for not only real-time electronic program guide for live broadcast programs but also refined auction of advertisement slots. More importantly, it demonstrates for the first time the feasibility of using Twitter for real-time social and physical event detection for ubiquitous computing.

## 1. Introduction
The global human population can be regarded as geographically distributed, multimodal sensors. However, only a specially tasked fraction of us are traditionally used in that way, e.g., journalists, explorers, and spies. World-Wide Web has made a significant fraction of us publishers who may report what we see and feel through blogs, forums, product reviews, and social networking sites. Such publications can be considered as the "readings" of the human sensors that contain information about the physical world that we sense.

Twitter provides a unique and novel venue of publishing: it has over 200 million active users around the globe; tweets are brief, limited to 140 characters, an ideal way for people to publish spontaneously. As a result, Twitter has the short delays in reflecting what its users perceive, compared to other venues such as blogs and product reviews. While many have demonstrated Twitter can provide insights into major social and physical events like earthquakes [1], celebrity deaths [2], and presidential elections [3], in this work, we answer a much tougher question: *how good is Twitter at real-time sensing for less significant but more frequent events such as what happen in a sport game*? It is our belief that insights gained from answering this question will fuel novel ubiquitous service innovations that leverage humans as sensors of the physical world.

Toward answering the above question, we report our experience with using Twitter to monitor the US National Football League (NFL) games. By analyzing tweets collected during the game time, we seek to recognize major game events as soon as they happen. The recognized events can be used to implement a better electronic program guide for live broadcast programs, which can provide more personalized, pertinent programs, or to provide a better pricing mechanism for selling advertisement slots [4], typically by sensing instantaneous popularity of a segment.

There are multiple challenges toward game event recognition using Twitter. First, we must separate the tweets related to a particular game from over 800 tweets generated every second. Although Twitter's #hashtags can be used to indicate the topics of tweets, only a small fraction (11%) of tweets contain hashtags [5]. Second, we must be able to distinguish which game a tweet is referencing when many games are played simultaneously, e.g., up to 10 during the 2010-2011 NFL regular season. Third, we must distinguish diverse and frequent events from sports games in comparison to singular events like earthquakes [1], celebrity deaths [2], and presidential elections [3]. Finally, this analysis has to be done in real-time with as short delay as possible. For example, to be useful for advanced advertising auctions, the event has to be recognized within tens of seconds in order to auction and display a customized advertisement.

To tackle these challenges, we investigated multiple methods for analyzing real-time Twitter data and found a simple lexicon-based method effective. We provide a two-stage solution using the Twitter Streaming API. The solution employs an adaptive sliding window to detect an event based on post rate change and then recognizes the event using lexicon-based content analysis. This solution is extremely efficient because it only does an in depth analysis of tweets after an event is detected. The adaptive sliding window design allows the accurate detection of events within 40 seconds after an event takes place.

Second, we report the first public study of real-time issues using Twitter for event recognition. We show that on average it takes 17 seconds for a Twitter user to report a game event. Surprisingly, we find that tweets from mobile devices are consistently several seconds slower than non-mobile tweets. We also find that Twitter's Streaming API has a delay between one second for tweets with popular keywords and hashtags, and about 30 seconds for those with uncommon tags.

Finally, we report our experience in a real-field deployment of the two-stage solution as a web service hosted on Amazon EC2 [6]. The service performed real-time event recognition for the last 101 of the 2010-2011 NFL games based on 22 million tweets from 4.7 million users. Through a web site, the service visualized the results and provided a popularity "thermometer" for games that were played at the same time. In this paper, we will show that the efficient, two-stage solution worked very well for all games but Super Bowl 2011, achieving 90% accuracy in recognizing major game events with a delay of only 40 seconds. However it failed catastrophically for Super Bowl 2011 due to an undocumented Twitter restriction. We will show that an improved solution that unifies the detection and recognition stages can not only deal with Super Bowl 2011 but also retain a high accuracy for early games under the Twitter restriction.

Although we focus on NFL games in this work, most of the techniques can be readily applied to many other sports games that have a similarly sized fan population and have similar frequencies of major events, e.g., soccer and baseball. More importantly, our success with NFL is a strong demonstration of the feasibility of Twitter as a real-time reading of the human sensors to detect social and physical events that can happen as diverse and frequent as sports events.

## 2. Related Work

While our work is the first to detect sports game events using Twitter, many have used Twitter to detect other social and physical events. Sakaki *et al* [1] and Qu *et al* [7] investigated the earthquakes detection. Vieweg *et al* [8] studied the grassfire and floods, on microblogs. TwitterStand [2] identifies current news topics and clusters the corresponding tweets into news stories. None of the above can detect the targeted event in seconds, which is key to fast-paced sports events. For example, the authors of [1] were able to detect an earthquake from Twitter only hours after it happened.

Several concurrent projects also study tweets about sports games, however they do not provide real-time event detection. Hannon *et al* [9] used post rate of tweets to produce video highlights of the World Cup off-line. They did not recognize game events nor did they produce highlights in real-time. Chakrabarti and Punera [10] assumed that a game event is already recognized and focused on describing the event using Hidden Markov Models trained with tweets collected from events happened in the past. Therefore, our focus on real-time event recognition is complementary, and addresses a more difficult and fundamental problem.

Event recognition for sports games has been studied by the video analysis research community. For example, Ekin et al [11] employed visual features analysis to summarize soccer videos. Rui et al [12] utilized speech detection techniques to extract events in baseball games. Zhang and Chang [13] utilized closed captioning text for baseball video event detection and summarization. Furthermore, multimodal approaches have been studied for video summarization in online presentation [14] and sports games [15, 16]. Petridis et al [17] and Xu et al [18] used MPEG-7 and webcast text to extract sports events. Availability is the primary problem to utilize external knowledge. Compared to these approaches, our Twitter-based approach enjoys several unique strengths. First, the video content and text information leveraged by the above work is not always available, especially in real-time. For example, NFL games include the closed captioning text in the video but do not offer webcast text. Moreover, video-based solutions are network and compute-intensive compared to twitter analysis, especially when multiple games are played at the same time, e.g., 10 during the NFL regular season. Most importantly, our twitter-based approach can be readily extended to recognize social and physical events beyond sports games as long as these events are witnessed by a large number of Twitter users.

## 3. Data collection

To study solutions to the challenges discussed above, we collect massive amount of Twitter data for a number of popular sports games in North America.

### 3.1 Twitter APIs

To achieve our goal of real-time analysis for event recognition, we retrieve as many relevant tweets as fast as possible and from as many users as possible. Twitter provides three application programming interfaces (APIs).

The Representational State Transfer (REST) API [19] allows developers to access core Twitter data stored in the main database which contains all the tweets. Through the REST API, developers can retrieve Twitter data including user information and chronological tweets. For example, the *home timeline* includes the 20 most recent tweets on a user's home page; the *public timeline* returns the 20 most recent tweets in every 60 seconds. These limitations make the REST API not particularly suitable for real-time tweet collection, the REST API is best for collecting a large number of tweets from specific user IDs off-line. In our study, we used it to collect tweets from NFL followers posted during the game for the 2010 Super Bowl off-line.

The *Search API* will return tweets that match a specified query; however it will only search a limited subset of tweets posted in past 7 days in the main database. The query parameters include time, location, language etc. Twitter limits the return results to 100 tweets per request. Although the Search API is able to collect tweets in real-time, one cannot control the topic of the returned tweets. Twitter limits the request rate to the REST and Search API to 150 per hour by default. It previously allowed up to 20,000 search requests per hour from white-listed IPs (about six requests per second), but, unfortunately, Twitter no longer grants white-listing requests since Feb 2011 [20]. This limitation makes the REST and Search APIs unsuitable for real-time event detection.

The *Streaming* API [21] offers near real-time access to Tweets in *sampled* and *filtered* forms. The *filtered* method returns public tweets that match one or more filter predicates, including *follow*, *track*, and *location*, which correspond to user ID, keyword and location, respectively. Twitter applies a User Quality Filter to remove low quality tweets such as spams from the Streaming API. The quality of service of the Streaming API is best-effort, unordered and generally at-least-once; the latency from tweet creation to delivery on the API is usually within one second [22]. However, reasonably focused track and location predicates will return all occurrences in the full stream of public statuses. Overly broad predicates will cause the output to be periodically limited [23]. Our experience shows that the Streaming API is better than either the REST or Search API for real-time game event recognition for three reasons: all the tweets returned are up to date; there is no rate limit; and the track filter predicate allows us to collect tweets that are related to the game of interest using keywords. Although there is no explicit rate limit, we cannot obtain all public tweets and we will report our observation of an undocumented restriction in the Streaming API.

```
Events Detection
1. At each second, initialize window size as 10
   seconds;
2. Post rate ratio = (post rate in the first
   half) / (post rate in the second half of
   slide window);
3. If (post rate ratio < threshold)
       Increase window size until 60 seconds;
       Go to step 2;
   Else
       Proceed to event recognition;
Event Recognition
1. Pre-processing
2. Compute the post rate of pre-defined event
   keywords in the second half of the window;
3. If (pre-defined event keyword appears >
   threshold)
       Recognize the event;
```

**Figure 1: Two-stage solution with event detection and recognition**

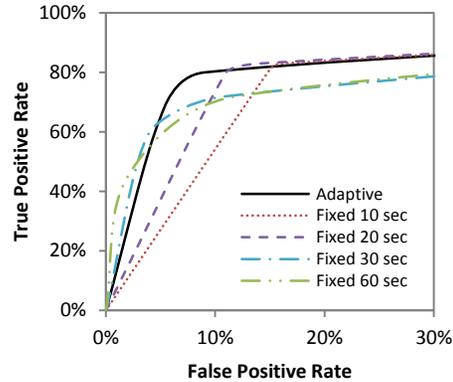

**Figure 2: The RoC curves for events detection with different window sizes.**

## 3.2 Targeted Games

We use live broadcast the US National Football League (NFL) games as benchmarks. We collected tweets from the 2010 Super Bowl and over 100 games in the 2010 to 2011 season including the 2011 Super Bowl. First, for the 2010 Super Bowl, we collected the tweets posted during the game, by followers of the NFL twitter account, or simply *NFL followers*, using the REST API. Overall, we collected over half a million tweets from 45,000 NFL followers. Although these tweets were collected off-line, they helped us gain insights into the keywords for collecting tweets in real-time with the Streaming API.

For the 2010-2011 season NFL games, we collected tweets during game time using the Streaming API and game keywords identified from the 2010 Super Bowl. We collected the tweets and their metadata such as tweet source, created time, location, and device. These tweets were analyzed for event recognition in real-time through a web service described below. For the regular season games and playoffs, we collected more than 19 million game-related tweets over a period of 9 weeks including 100 games, from 3.5 million users. We collected about 1 million game-related tweets from over half a million users for 2011 Super Bowl. The evaluation of our solutions was performed in real-time when a game was ongoing and was repeated with trace-based emulation off-line if necessary.

## 3.3 Lexicon-based Game Tweets Separation

We next provide our rationale behind the keywords used to retrieve game-related tweets for real-time analysis. Here we use the tweets retrieved with the REST API from NFL followers posted during the 2010 Super Bowl. When performing real-time analysis of tweets for a game of interest, it is very important to focus on tweets that are actually talking about the game, not only because tweets unrelated to the game will interfere with the analysis but also because Twitter limits the rate tweets can be retrieved (Yes, even for the Streaming API as we will see later).

We find that such keywords include game terminology and team names. To examine the relationship between the game and tweets posted during the game, we compute and rank the term frequencies of all words that appeared in the tweets posted during the game. After running a stemming algorithm to eliminate misspelling words, we find that the top 10 most frequent words are either game terminology or team names.

Are these keywords sufficient to extract game-related tweets? To answer this question, we randomly select 5% of the tweets, about 2,000, posted during the game by the NFL followers. We manually examined all of these tweets to determine if each of them was related to the game. Half of these tweets had at least one of the top 10 keywords. There were some tweets with incomplete sentences; and we treated the uncertain ones as unrelated. Using the manually classified set of tweets as the ground truth, we find that extraction by the top 10 keywords is surprisingly effective, achieving a false negative rate below 9% and a false positive rate below 5%.

Further, we found that the team names appear in over 60% of the game-related tweets. Therefore, we rely on the team names to collect data when multiple games take place at the same time and attribute these tweets to games based on the mentioned team names.

The performance of the lexicon-based heuristic can be further improved by examining the falsely identified tweets. The first major source of error is the foreign languages tweets using the keywords or Twitter hashtags. If these tweets are not considered, the false positive and false negative rates will be reduced to 5.2% and 2.8%, respectively. The second major source of error is misspelling because Twitter users can spell words wrong either deliberately or by mistake. By applying the spelling check engine and regular expression applications, we can reduce the false positive and false negative rate to 4% and 2%.

## 4. Event Recognition

In this section, we show that NFL events can be recognized by examining the post rate and analyzing the content of game-related tweets collected using the lexicon-based heuristic. We report a two-stage solution in which an event is first detected and then recognized as described by Figure 1. The two-stage structure helps reduce the computational load significantly because detection can be achieved without analyzing the content of tweets. We applied this solution to detect pre-defined football game events including touchdown, interception, fumble, and field goal for nearly a hundred NFL games in real-time. While this solution worked very well for games in regular season and playoffs, in Section EVALUATION, we will show how an undocumented Twitter constraint failed it for the 2011 Super Bowl and how it can be fixed.

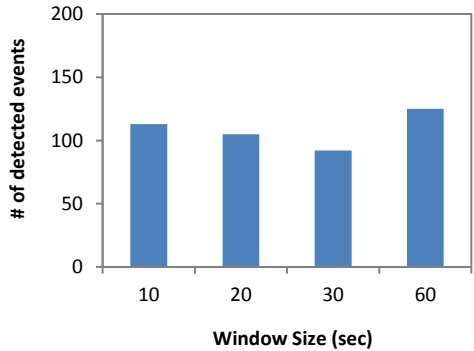

**Figure 3: The distribution of events detected with various window sizes**

## 4.1 Adaptive Rate-based Detection

By visualizing the post rate of game-related tweets during the game time, we observe prominent post rate spikes usually coincide with important events. Prior work [1, 2] has already shown that the increase in post rate is closely related to the significant events such as earthquake or celebrity death. Although sports games like the NFL games have many watchers in real-time, they are not as significant as the breaking news. More importantly, sports games usually produce many events in a short time span.

### 4.1.1 Adaptive Sliding Window-based Detection

Based on our observation, a simple method to detect an event would calculate the post rate increase as the ratio of the post rate in the second half of a sliding time window to that in the first half. The size of the window will have a significant impact on the tradeoff between the delay and accuracy of event detection. A shorter window will lead to a smaller delay but may have a poor performance when the post rate is low and, therefore, there are not many tweets posted in the time window.

To achieve the best tradeoff, we devise a solution that selects the window size and the threshold for post rate increase adaptively. The sliding window has a variable size of 10, 20, 30, or 60 seconds; and each window is divided temporally into two sub-windows containing the tweets in the first and second half of the window. At every second, the program will start from the shortest window, 10 seconds, to examine the post rate between the two halves. If the post rate ratio exceeds the threshold, the program will proceed to recognize events, otherwise the window size will increment. The dynamic threshold contains two parts, the post rate ratio and the average number of tweets. The post rate ratio is the number of tweets in the second half window to the number of those in the first half. The value is set to 1.7 in this case; that means the post rate in the second half window needs to be at least 1.7 times of the post rate in the first half to proceed. The average number of tweets serves as a relatively stable threshold to filter out the minor increases because the main spikes appear all above the average.

### 4.1.2 Detection Performance

Like any binary classifier, the detection stage can make two types of errors: reporting an event when nothing happens (False positive) and reporting nothing when an event happens (False Rejection). Using all 435 events, 163 touchdowns, 68 interceptions, 91 field goals, and 113 fumbles, happened in 32 games in the Week 16, 17 and playoffs of the 2010-2011 NFL season, we illustrate the effectiveness of our adaptive method.

Figure 2 shows the Receiver Operating Characteristics (RoC) Curves for our adaptive methods and those based on a fixed sliding window. We compared the performance of the adaptive window approach with the fixed window approaches. From the results in the RoC curve, the adaptive window size outperforms the fixed window size.

Figure 3 shows the number of events detected in each window size; half of events can be detected using the window size less than 20 seconds and more than 2/3 of events can be detected using the window size less than 30 seconds. Since we halve the window to detect events, the delay of the system is half of the window size. As a result, this method will introduce less than 10 second delay for half of the events.

## 4.2 Lexicon-based Recognition

Once an event is detected, we employ a lexicon-based recognition method using the post rate of events keywords. Our results show that major events, i.e. touchdown, interception, in NFL can be accurately recognized.

Before we extract the event keywords, we first remove URLs, @username (replies to users), emoticons, and punctuations. Then we utilize the process method described in previous sections to remove non-English words, stop words and stem the remaining words. We calculate and rank the post rate of the events keywords. We select and determine the event with highest post rate.

The lexicon-based recognition also helps suppress false alarms from the event-detection stage. If a random event rather than the major event causes the general post rate increase, the lexicon based recognition will reject the random event.

## 5. Real-Time Considerations

Delay or latency is an important consideration for any sensing apparatus. In this section, we investigate it for the sake of using Twitter as a real-time reading of the human sensors. Many applications require game events and game popularity information to be detected in real-time. The information provided to end users or advertisers must be on time. Otherwise, the information is meaningless when people are no longer interested in the event. For example, the EPG needs the game information in real time, advertising campaign is operated near real time. Therefore, we analyze the delay of our Twitter-based solutions in this section.

Three sources contribute to the delay of our Twitter-based solution. The delay introduced by Twitter users is the period between people perceiving a social or physical event and posting the tweet. This *human delay* is mainly determined by how fast twitter users perceive the event, how fast they react to it, and how fast they type the tweet. Twitter itself also introduces delay in providing tweets through its API; *the Twitter delay* may be affected by Twitter workload, the user quality filter process or the Twitter's index mechanism. Finally, our data collection and analysis also introduce a delay or *analysis delay*.

To study these delays, we must be able to know the following important time stamps: when an event takes place, when a related tweet is submitted to Twitter by user, when the tweet is retrieved through the Stream API, and when the event is recognized. While the last two time stamps are trivial to obtain, the first two requires some extra work. First, oddly all media, e.g., ESPN channel, NFL website and sports newspapers, record the game events in the

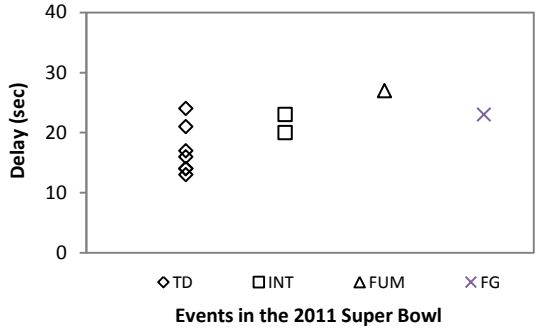

Figure 4: The distribution of human delay of various events in the 2011 Super Bowl

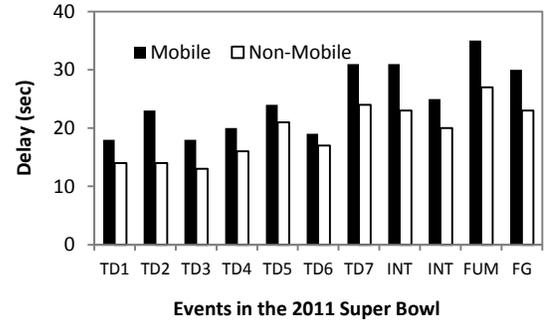

Figure 5: The human delay of tweets posted from mobile and non-mobile when events happen.

*game time*, the time corresponding to the game progress. Moreover, the NFL game rewind is commercial free, making the time a few minutes apart from the absolute time. Therefore, we chose to video-tape the live broadcasted 2011 Super Bowl and record the event time manually as the ground truth. We recorded 7 touchdowns, 2 interceptions, 1 fumble, and 1 field goal. Secondly, Twitter time stamps a tweet when it is received. We call this time the *Twitter time*. To accurately estimate the human delay, we must know if there is a significant delay between when a tweet is submitted and when it is time-stamped by Twitter. We found the delay is negligible, i.e., one second or shorter. We can prove that the Twitter time represents the time when Twitter receives tweets in the following experiment. We run the data collection program to track the keyword "Twittertime" which is created by us. Then we post tweets containing the current time (in the precision of second) in our system with the keyword. The collected tweets contain the enclosed Twitter time either the same or only 1 second later than the system time specified in the tweet.

## 5.1 Human Delay

We estimate the human delay as the time difference between when a game event happens and when Twitter time stamps a tweet talking about the event. We found that the human delay is as short as 13 seconds and people using mobile devices tweet are even slower than people using non-mobile devices.

We extract the tweets posted after the event and contain the event keyword. Then we read the tweets to select the tweet that is the first discussing the event in a narrative statement rather than prediction or anticipation. People predict events, discuss about the past events which introduce some noise but in a low post rate and a low frequency, i.e. at most 2 in a second and usually null. Since two events of the same type do not usually happen in a short interval (several minutes), we assume the tweets that mentioned a happened event posted several minutes after an event of the same type are discussing about the event that just happened. Our results as summarized by Figure 4 show that the average human delay is 17 seconds. The shortest delay is only 13 seconds and the longest is 27 seconds. Interestingly, touchdowns saw shorter human delays than other less significant ones, indicating Twitter users post faster for more significant events.

It must be noted that there is a short delay in broadcasting live materials used to prevent profanity, bloopers, violence, or other undesirable material from making it to air. This delay may vary in different locations and is approximately 7 to 12 seconds. As a result, tweets by game watchers from the stadium should be posted earlier than those from homes. However, we were not able to collect a large number of tweets that can be identified as from the stadium because only a small fraction of Twitter users allow their location to be included in their tweets.

### 5.1.1 Mobile Tweets are Slower

We had expected tweets from mobile devices to have a shorter delay because the effort to switch from game watching to tweeting seems to be lower on mobile devices than on PC or laptops. Our results surprisingly show the opposite. By examining the source clients, we observed that nearly 40% of game-related tweets were from Twitter clients on recognizable mobile devices, i.e. iPhone, BlackBerry, Android, txt, mobile, HTC, MOTO, and iPad. The actual percentage of tweets from mobile devices should be more than 40% because there are tweets posted from the clients that are both available to mobile and non-mobile devices. In this study, we only consider the tweets from the recognizable mobile devices.

We inspect 7 touchdowns, 2 interceptions, 1 fumble, and 1 field goal in the Super Bowl. The results illustrate that the non-mobile users react 3 to 5 seconds faster in all the 11 events, as shown in Figure 5. The number of tweets posted from non-mobile devices increases faster as well. Take the $1^{st}$ touchdown for example, the poste rate from non-mobile devices increases from null to the maximum (27 per second) in 10 seconds whereas that from mobile devices starts 3 second later and takes 36 seconds to reach the maximum (23 per seconds).

The possible reasons that cause the results include typing speed and device/network delay. Although mobile devices, mainly smartphones, are more portable and convenient, the typing speed on mobile devices is still lower than that on PC or laptops. On the other hand, the mobile devices may suffer from the network condition [24]. Consequently, mobile users spend more time in the *human delay*.

## 5.2 Twitter Delay

Twitter Streaming API allows keyword-based retrieval in real time. We are interested in how much delay Twitter introduced to its Streaming API since there is no published study regarding it. We calculate the Twitter delay as the difference between the Twitter timestamp of a tweet and the time we obtain the same tweet from Twitter.

We make two observations regarding the Twitter delay. The delay is about 30 seconds for tweets only contain custom, random keywords and about one second for tweets with Twitter promoted, widely used keywords, e.g., Twitter promoted #sb45 during the 2011 Super Bowl. Second, the delay is fairly independent of the post rate.

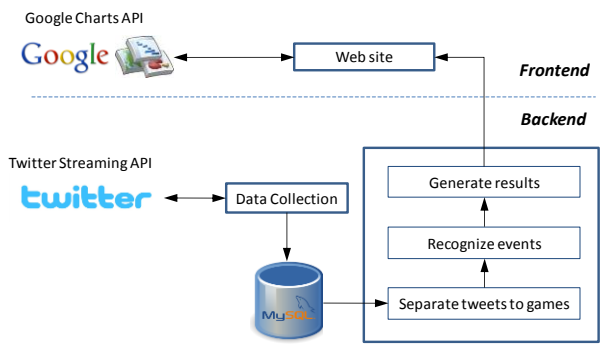

Figure 6: Architecture of our web service implementation

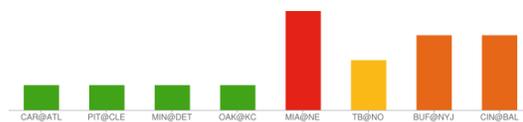

(a) Color-coded thermometer for concurrent games

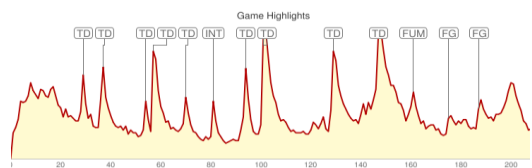

(b) Excitement level and event recognition results

Figure 7: Real-time visualization from our website

Our observations are based on two measurement studies. The first measurement is similar to that used to measure the human delay. Tweets that only contain the system time and the keyword "Twittertime" are posted and retrieved through the Streaming API to calculate the Twitter delay. 10 tweets were randomly posted in 2 hours, once in a weekday afternoon without special event, and the other during the NCAA Men's Basketball Final. The Twitter delay is consistently about 30 seconds.

The second measurement was performed during the NCAA Men's Basketball Final with a keyword promoted by Twitter, *#MM2011* meaning March Madness 2011, and the widely used team names, Butler and UConn, to collect data. Like the NFL playoffs, the post rate was high during the game (20 per second). We found the Twitter delay is consistent about one second, much shorter than the 30 seconds we observed for retrieval with custom keywords. Even the team names, which are widely used but not promoted by Twitter, when used as the keyword, return tweets in about one second, which is consistent with Twitter's own claim [22].

It is surprising that the choice of keywords makes such a huge difference in Twitter delay: 30 seconds vs. 1 second. Without information from inside Twitter, we have to speculate the reason. Twitter maintains its Streaming API result quality by applying the status quality metrics, in some instances, in combination with other metrics, to filter the tweets [23]. The goal is to eliminate spams, inappropriate, or repetitious tweets. Therefore, the tweets contains rarely used keywords may take longer to pass the process. In addition, developers in the Twitter development talk, an online forum in Google groups [25], mentioned that Twitter may index interesting tweets to improve the search speed. Since the Streaming API with filter predicates, i.e., follow, track, or locations, involves search, we speculate such indexing helps significantly reduce the Twitter delay for Twitter promoted keywords and popular keywords such as team names[26, 27].

We also observe that the post rate does not impact the Twitter delay noticeably by comparing the delay during both low and high traffic time, i.e., during the March Madness 2011.

### 5.3 Analysis Delay

The data collection and processing to recognize events also introduce delay. We have implemented our solutions in a way to maximize the parallelism of data analysis as will be described in the next section. Here we report the measurement of the delay in our implementation.

We estimate that the analysis is less than 20 seconds on average. The delay introduced by the detection stage is bounded by the used window size, 10 to 60 seconds or a delay of 5-30 seconds. As we have shown in Figure 3, the average value is 15 seconds. The second stage introduces delay purely from its computing tasks, which take less than 5 seconds on an average server. Because the human delay is about 20 seconds and the Twitter delay is about one second, our solution recognizes a game event around 40 seconds on average after the event takes place. The sliding window used in event detection contributes a lot to the overall delay. Interestingly, the higher the post rate, the shorter window will be used and, therefore, introduce a shorter delay. For major events such as touchdowns, the overall delays can be as short as 30 seconds. To appreciate the delay achieved by our Twitter analysis, we find that the ESPN web page has 90 seconds lag in updating the latest score changes.

## 6. Implementation, Evaluation, and Improvement

We have implemented the two-stage solution described above as a real-time web service that visualizes event recognition results through a website, sportsense.us, throughout the 2010-2011 NFL season [28]. We next describe the implementation and report our experience with its performance. In particular, we show how an undocumented Twitter restriction failed the otherwise successful service during the 2011 Super Bowl and offer an effective fix to it.

### 6.1 Web Service Realization

The implementation is coded in PHP and consists of the backend for data collection and analysis and the frontend for web visualization, as illustrated by Figure 6. The service was originally hosted on a lab server and later moved to Amazon EC2 which is a reliable, reasonably priced cloud computing platform.

The backend consists of two parallel modules and a MySQL database. The data collection module collects game-related tweets through Twitter Streaming API as described in Section DATA COLLECTION. Collected tweets are saved in the MySQL database. The event recognition module will retrieve tweets from database, separate tweets to games, recognize events, and generate the results in Google Chart format.

As analyzed above, the backend can introduce many seconds of delay to event recognition. To minimize this delay, we created multiple threads to maximize the parallelism of data analysis. The data collection module employs one thread to retrieve tweets from Twitter, decode and save the tweets into the MySQL database.

**Table 1: The confusion matrix of event recognition by the two-stage solution**

|  |  | Actual | | | | |
|---|---|---|---|---|---|---|
|  |  | TD | INT | FG | FUM | NULL |
| **Recognized** | TD | 146 |  |  |  | 18 |
|  | INT |  | 55 |  |  | 20 |
|  | FG |  |  | 77 |  | 30 |
|  | FUM |  |  |  | 73 | 16 |
|  | NULL | 17 | 13 | 14 | 40 |  |

The event recognition runs as another thread that retrieves data from the database and analyze them for event recognition.

The frontend visualizes the analysis results using Google Charts API through a website. For ongoing games, the website shows a color-coded bar chart for the "hotness" of all concurrent games according to the post rate of tweets related to each game, as shown in Figure 7 (a). For each game, the website provides a line chart that draws post rate of tweets related to the game and denotes recognized events, as shown in Figure 7 (b). To update the website in real-time, the website employs an embedded Javascript to pull the results from the backend every two seconds. The line charts and recognized events for past games can be retrieved from the website by either team name or game schedule.

We choose Amazon EC2 to host the web services for its reliability and the full control to the virtual machine which Amazon calls an *instance*. We are using the small (default) standard instance which provides 1.7GB of memory, 1 virtual core and 160GB of local instance storage [6]. This instance is charged on-demand for $0.085 per hour for Linux/Unix. The data transfer cost is $0.1 per GB for data transfer IN and $0-0.08 per GB for data transfer OUT depending on the volume. As our system is always on and generates less than 5GB transfer in data per month, the operational cost is about $60 per month.

## 6.2 Evaluation Results

The web service has been active since Week 8 of the 2010-2011 NFL season. Because we used it to improve the solution in the first weeks, the final solution presented in this paper was implemented in time only for the last two weeks (27 games in Week 16-17), playoffs (5 games), and the Super Bowl. Using the game analysis from the NFL website as the ground truth, we are able to evaluate the two-stage solution. Table 1 summarizes the confusion matrix of our event recognition. It shows not only how many events have been correctly recognized but also how many are missed (Null). Note that four events are targeted: touchdown (TD), interception (INT), field goal (FG), and fumble (FUM). A special "event" NULL is used to represent eventless game time. We make the following observations. First, the only misrecognitions are false alarms and false positive; once an event is detected, it is recognized reliably. This indicates the strength of the recognition stage but suggests potential improvement is necessary for the detection stage. Second, our solution works much better for more significant events. For example, it does very well for touchdowns (TD), the most significant event in a football game: 146 out of 163 touchdowns are recognized with 18 false positives (90% true positive rate). The performance is, however, much worse for the least significant event, fumble (FUM), with 64% true positive rate. The system performs unsatisfactory for fumble because people do not tweet about it especially when fumble is recovered by the fumbler so that it cannot cause the significant change on the post rate. The high false positive rate of field goal is because people also use this term for the point after touchdown (PAT). While the performance with touchdowns is very good, significant improvement is required for minor events.

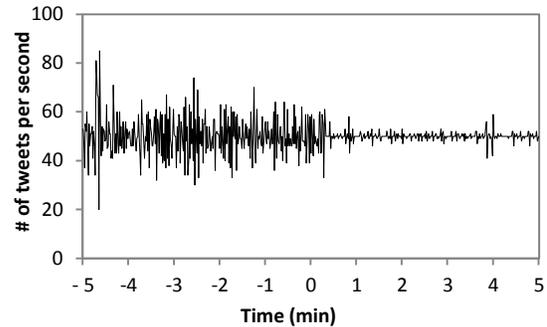

**Figure 8: The second-wise post rate of collected tweets for five minutes before and after the first touchdown in the 2011 Super Bowl**

## 6.3 A Case of Failure and Its Solution

While our system performed consistently well for all the games in the final weeks and playoffs, it catastrophically failed to recognize any event during the 2011 Super Bowl. In the rest of this section, we present results from our investigation of the failure and provide enhanced solutions to cope with similar situations.

Since we recorded all the tweets for the games, we were able to investigate what happened. Our investigation revealed that the failure was caused by an implicit throughput limit by Twitter for its Streaming API: about 50 tweets per second. Twitter API documentation only explicitly states two other limits: first, "reasonably focused track and location predicates will return all occurrences in the full Firehose stream of public statuses" and "broad predicates will produce limited streams that will tend to be a subset of the statuses/sample streams" (1% sampling rate).

We discovered the undocumented throughout limit of the Streaming API by analyzing the tweets collected during the Super bowl. We notice that the post rate does not increase considerably as usual when events happen. We therefore zoom in to examine the number of tweets collected in each second from 5 minute before to 5 minutes after the event. To our surprise, we find that the throughput is floating around 50 tweets per second and does not increase when an event happens. In particular, the average number of tweets collected per second is 50.1 and the limit becomes stricter immediately after the event because the standard deviation reduces dramatically from 8.4 to 1.9, as shown in Figure 8. Had not been this limit, we should have collected tweets at much higher rate when an event happens because, according to Twitter [29], the overall post rate doubled during major Super Bowl events.

The event detection stage failed to detect any rises in the post rate of tweets collected from the Streaming API because the true post rate during the Super Bowl was almost always very high, keeping our data collection at the throughput limit of 50 tweets per second. In contrast, in the regular season and playoffs, the rate limit was never reached so that the event detection stage did see the post rate rises due to events.

### 6.3.1 Solutions

While Twitter probably supports a much higher rate for paid users, a solution that works under the limit is highly desirable. We revise the original two-stage solution to be independent to the post rate change without compromising the accuracy or real-time performance. Our key observation is: the post rate of tweets with events keywords remains well below the 50 per second and its increase on the corresponding event is obvious even under extremely high post rate.

We utilize this finding to merge the event detection and recognition stages into one. The new *unified* solution tracks events keywords and detects, recognizes events at the same time based on the keywords post rate. Given a tweet, we first search the events keywords, i.e., touchdown, fumble, field goal, and interception. If the event keyword appears, we count its post rate in this second. Then we apply the same adaptive sliding window size-based events detection approach discussed in the EVENT RECOGNITION Section. The difference is that we feed in the post rate of tweets with event keywords instead of the post rate of all game-related tweets.

Because we have recorded all the tweets from the NFL season including the Super Bowl, we are able to evaluate the unified solution. The unified solution is able to recognize all the events in the Super Bowl game without false positives and also maintain the performance for the games in the regular season and playoffs (Compare Table 1 and Table 2). The solution also incurs delays similar to the original two-stage solution.

The unified solution, however, incurs higher computation intensity because it examines every game-related tweet for event keywords. In contrast, the original two-stage one only has to do so when an event is detected by examining tweet post rate. Our experiments, on the other hand, show that the additional computing incurred by the unified solution introduces almost unnoticeable difference when executed by powerful servers in the cloud, e.g. Amazon EC2.

## 7. Conclusion

In this work, we pushed the limit of Twitter as a real-time reading of the human sensors. We demonstrated that moderately frequent and diverse social and physical events like those of NFL games can be fairly reliably recognized within 40 seconds using tweets properly collected in real-time. We hope this new capability will inspire more applications in the ubiquitous computing community.

Our results also suggest that more research is needed to further improve the performance of Twitter-based event recognition, in particular for less significant events. Moreover, our work is limited to events for which keywords can be *predetermined* like the NFL game events. What would be more useful and challenging is to recognize events that are not anticipated and, therefore, do not have keywords defined beforehand. Leveraging the work reported here, we are currently actively pursuing this goal.

**Table 2: The confusion matrix of event recognition by the unified solution**

| | | Actual | | | | |
|---|---|---|---|---|---|---|
| | | TD | INT | FG | FUM | NULL |
| Recognized | TD | 151 | | | | 15 |
| | INT | | 51 | | | 10 |
| | FG | | | 61 | | 8 |
| | FUM | | | | 84 | 20 |
| | NULL | 12 | 17 | 30 | 29 | |